\documentclass[a4paper,11pt]{article}
\pdfoutput=1 

\usepackage{jheppub} 

\usepackage[T1]{fontenc} 

\subheader{CERN-PH-TH-2014-199, DFPD-2014/TH/16}

\title{\boldmath On sgoldstino-less supergravity models of inflation}
\author[a]{Gianguido Dall'Agata}
\author[a,b]{and Fabio Zwirner}
\affiliation[a]{Dipartimento di Fisica e Astronomia `G.~Galilei', Universit\`a di Padova 
\\ 
and INFN, Sezione di Padova, Via Marzolo 8, I-35131 Padova, Italy}
\affiliation[b]{Theory Unit, Physics Department, CERN, CH-1211 Gen\`eve 23, Switzerland}
\emailAdd{gianguido.dallagata@pd.infn.it}
\emailAdd{fabio.zwirner@pd.infn.it}

\abstract{
We go a step further in the search for  a consistent and realistic supergravity model of large-field inflation by building a class of models with the following features: during slow-roll, all the scalar fields other than the inflaton are frozen by large inflaton-dependent masses or removed from the spectrum; at the end of inflation, supersymmetry is spontaneously broken with naturally vanishing classical vacuum energy. 
We achieve this by combining some geometrical properties of the K\"ahler potential with the consistent use of a single nilpotent chiral superfield, in one-to-one correspondence with the supersymmetry-breaking direction in field space at the vacuum.
}
\def\simlt{\mathrel{\lower2.5pt\vbox{\lineskip=0pt\baselineskip=0pt
           \hbox{$<$}\hbox{$\sim$}}}}
\def\simgt{\mathrel{\lower2.5pt\vbox{\lineskip=0pt\baselineskip=0pt
           \hbox{$>$}\hbox{$\sim$}}}}
\begin{document} 
\maketitle
\flushbottom

\section{Introduction}
\label{sec:intro}

If we ignore the big unsolved problems of the quantum theories of the fundamental interactions, such as the cosmological constant problem and the naturalness problem of the weak scale, and we limit ourselves to small perturbations around classical effective actions, we are not very far from producing realistic and consistent supergravity models for (large-field) inflation. 
However, even in this limited context there are still several problems to address, both from the point of view of consistency and from the point of view of realism. 

In the rest of this introduction we review three main problems that have to be faced in building supergravity models for large-field inflation:
slow-roll conditions, to realise inflation;
supersymmetry breaking in flat space at the end of inflation, to connect with low-energy physics;
effective single-field inflation, the simplest option and a sufficient condition to avoid unacceptably large isocurvature fluctuations.
In Section~2 we first describe two special classes of supergravity models: no-scale models, whose geometrical properties allow for spontaneous supersymmetry breaking with naturally vanishing classical vacuum energy, at the price of a classically massless complex scalar; models with a nilpotent chiral superfield, where the complex scalar partner of the goldstino is consistently removed from the spectrum by imposing a supersymmetric quadratic constraint. 
We then show how to combine the good features of these two classes to obtain classical supergravity models with broken supersymmetry, naturally vanishing vacuum energy and no massless scalars in the spectrum. 
In Section~3 we critically review the current state of the art in building supergravity models of large-field inflation with the use of nilpotent chiral superfields. 
Then, in Section~4, we present our main new results, which generalise the single-field toy model of Section~2 to inflationary models with two chiral superfields. 
We build a class of models where: during slow-roll, all the scalar fields other than the inflaton are either frozen by large inflaton-dependent masses or removed from the spectrum; the nilpotency constraint can be solved consistently on the full field space; at the end of inflation, supersymmetry is spontaneously broken with naturally vanishing classical vacuum energy, and the supersymmetry-breaking direction in field space is exactly aligned with the nilpotent multiplet. 
We conclude in Section~5 with some comments on how to extend our results to models with more than two chiral superfields and on the relation between perturbative unitarity during inflation and the scale of supersymmetry breaking in our models. 
On the first issue, we find that the matter scalars in the visible sector can be easily frozen by large inflaton-dependent mass terms during inflation. 
On the second issue, we find that, although we might expect a stringent lower bound on the gravitino mass, the inflaton dependence of the auxiliary field of the nilpotent multiplet ensures compatibility with perturbative unitarity  during inflation also in the case of weak-scale supersymmetry. 

We begin by briefly recalling the structure of the supergravity potential. 
For the purposes of the present paper, we can neglect gauge interactions and couple $N=1$ supergravity only to chiral multiplets $Z^I \sim (z^I, \psi^I,F^I)$. 
The theory is then defined by the K\"ahler potential $K(Z,\overline{Z})$ and by the superpotential $W(Z)$, conveniently combined in the real function
\begin{equation}
\label{gfunct}
{\cal G} (Z,\overline{Z}) = K(Z,\overline{Z}) + \log |W(Z)|^2 \, , 
\end{equation}
invariant under the K\"ahler transformations
\begin{equation}
\label{ktransf}
K(Z,\overline{Z}) \rightarrow K(Z,\overline{Z}) + \Lambda(Z) + \overline{\Lambda} ( \overline{Z}) \, , 
\quad \quad
W(Z)  \rightarrow  e^{- \Lambda(Z)} \, W(Z) \, , 
\end{equation}
where $\Lambda(Z)$ is an analytic function of the chiral superfields.
The scalar potential reads
\begin{equation}
\label{potential}
V  = F^I \, F_I - 3 \, |W|^2 \, e^K = e^{\cal G} \, \left( {\cal G}^I \, {\cal G}_I - 3 \right)  \, , 
\end{equation}
where 
\begin{equation}
\label{auxI}
F_I = e^{K/2} \, D_I \, W \, , 
\qquad
F^I = e^{K/2} \, K^{I \bar{J}} \, \overline{D}_{\bar{J}} \overline{W} \, ,
\qquad
D_I W = W_I + W K_I \, .
\end{equation}
As usual in the supergravity literature, lower indices denote derivatives as in $W_I = \partial_I W = \partial W / \partial Z^I$, bars denote complex conjugation, $K_{\bar{J}  I} = \overline{\partial}_{\bar{J}} \partial_I K$ is the K\"ahler metric, indices are raised with the inverse K\"ahler metric $K^{I \bar{J}}$  ($K^{I \bar{J}} \, K_{\bar{J}  L} = \delta^I_L$), and we work in natural Planck units where $M_P^{-2} = 8 \pi G = 1$. 
The gravitino (squared) mass is $m_{3/2}^2 = 3 \, \langle e^{\cal G} \rangle$, Minkowski vacua with spontaneously broken supersymmetry correspond to 
\begin{equation}
\label{minkbro}
\langle F^I \, F_I \rangle = 3 \, \langle |W|^2 \, e^K \rangle \equiv 3 \, \langle e^{\cal G} \rangle \ne 0 \, ,
\end{equation}
while those with unbroken supersymmetry in flat space have 
\begin{equation}
\label{minkunb}
\langle F^I \rangle = 0 \quad (\forall I) \, , 
\qquad
\qquad
\langle W \rangle =0 \, . 
\end{equation}
We are now ready to complete this introduction by describing three plausible milestones in the path towards a realistic and consistent supergravity model for large-field inflation. 

\subsection{Slow-roll in supergravity}

In supergravity, the inflaton field $\varphi$ is usually identified with one of the two real scalar degrees of freedom of a chiral multiplet $\Phi$, the inflaton multiplet. 
A minimal requirement for model building is that the slow-roll conditions are satisfied in the inflaton direction. Assuming a canonical Einstein term for gravity and a canonical kinetic term for the inflaton,
\begin{equation}
\label{canonical}
S = \int d^4x \ \sqrt{-g} \ \left[ \frac{1}{2} \, R - \frac{1}{2} \, (\partial \varphi)^2 - V(\varphi) \right] \, , 
\end{equation}
and neglecting for a moment all the other scalar fields, the slow-roll conditions read
\begin{equation} 
\label{slowroll}
\epsilon = \frac{1}{2} \left( \frac{V^\prime}{V} \right)^2 \ll 1 \, , 
\qquad \qquad
\eta= \left| \frac{V^{\prime \prime}}{V} \right| \ll 1 \, , 
\end{equation}
where $V^\prime \equiv \partial V /\partial \varphi$ and $V^{\prime \prime} \equiv \partial^2 V /\partial \varphi^2$. 
Well-known examples of potentials for large-field inflation fulfilling (\ref{slowroll}), originally introduced in a non-supersymmetric context, are those associated with the simplest model of chaotic inflation \cite{linde-ch},
\begin{equation}
\label{linde}
V= \frac{1}{2} \ M^2 \, \varphi^2 \, ,
\qquad \qquad
M \sim 10^{-5} \, , 
\end{equation}
and with the dual version of the $R^2$ model by Starobinsky \cite{starob},
\begin{equation}
\label{staro}
V= V_0 \, \left( 1 - e^{- \sqrt{2/3} \, \varphi} \right)^2 \, ,
\qquad \qquad
V_0 \sim 10^{-9} \, .
\end{equation}

An initial difficulty in building supergravity models for large-field inflation fulfilling the slow-roll conditions (\ref{slowroll}) was the fact that the potential (\ref{potential}) is proportional to $e^K$, thus to $e^{\varphi^2}$ in models with canonical K\"ahler potential $K = |\Phi|^2$ for the inflaton multiplet, which makes $\eta$ of order one. 
It is by now well known that the way out \cite{yanag} is to impose a shift symmetry on the inflaton K\"ahler potential. 
It will not be restrictive to assume that
 \begin{equation}
 \label{infdec}
 \Phi = \phi + \sqrt{2} \, \theta \chi + \theta \theta F^\Phi \, , 
 \qquad \qquad
 \left( \phi = \frac{a + i \, \varphi}{\sqrt{2}} \right) \, , 
\end{equation}
so that the inflaton $\varphi$ is the imaginary part of the complex scalar $\phi$ in $\Phi$ and that the corresponding shift symmetry in the K\"ahler potential translates into $K=K(\Phi + \overline \Phi)$. 

\subsection{Minkowski vacuum with broken supersymmetry}

A second important requirement is that at the end of inflation, i.e. once the inflaton field has relaxed to its vacuum expectation value, supersymmetry should be spontaneously broken in an approximately flat background. 
The present vacuum energy density corresponds to $\langle V \rangle \simeq 10^{-120}$, a negligible value with respect to most scales of particle physics.  
Moreover, we have not observed any supersymmetric particle up to the TeV scale, and the supersymmetry-breaking masses $\widetilde{m} \simgt 10^{-15}$ are bounded from above by the supersymmetry-breaking scale $\langle F^I F_I \rangle^{1/4}$, thus we should ask that at the end of inflation $\langle F^I F_I \rangle > 10^{-60}$. 
The extraordinarily precise cancellation required for fulfilling simultaneously the two above constraints is the cosmological constant problem, manifest already at the classical level. 
Since the difficult problem is to explain the smallness of $\langle V \rangle$, not the fact that $\langle V \rangle \ne 0$, in this paper we will consider $\langle V \rangle =0$, i.e. a Minkowski vacuum. 
Holding the details for the following discussion, we mention here that a class of supergravity models whose geometrical properties lead to spontaneous symmetry breaking  with naturally vanishing classical vacuum energy are the so-called no-scale models \cite{noscale}.

A large fraction of the supergravity models of inflation proposed so far and fulfilling the slow-roll conditions (\ref{slowroll}) have supersymmetric Minkowski vacua, as in (\ref{minkunb}). 
This is often justified by the generic statement that the scale of inflation is typically much larger than the scale of supersymmetry breaking on the vacuum, thus supersymmetry breaking can be ignored in first approximation and treated as a perturbation. 
In our opinion, however, the validity of such a statement should not be given for granted in the absence of an explicit model, also in view of the considerations of the following subsection.

\subsection{Effective single-field inflation}

In the spectrum of a realistic supergravity model, the inflaton is not the only scalar field.
Indeed, there are many others, both in the `hidden' sector describing inflation and supersymmetry breaking and in the `observable' sector containing the Standard Model. 
To effectively realise the simplest single-field version of inflation, which guarantees the absence of unacceptably large isocurvature fluctuations, we can ask that during inflation all these extra scalar fields are stabilized to some definite background values by  large inflaton-dependent mass terms. 
Such condition is not so easy to fulfill, especially if we take into account all the scalar fields whose presence is required in a realistic model.  

An obvious presence is the second real scalar in the inflaton multiplet, $a$ in the notation of Eq.~(\ref{infdec}). 
A notable exception occurs in the models of D-term inflation with gauged shift symmetry  \cite{FKR,copernicus,fp}, where the scalar partner of the inflaton is removed from the spectrum by a supersymmetric Higgs effect.  
However, these models have supersymmetric Minkowski vacua and cannot be easily perturbed to break supersymmetry in a realistic way~\footnote{For an interesting attempt to describe inflation and supersymmetry breaking in higher-curvature pure supergravity, although with a fine-tuning to cancel the classical vacuum energy, see \cite{DFKRU}.}, since one of their crucial features is the vanishing of the superpotential on the vacuum.

If supersymmetry is spontaneously broken in flat space, there is always a chiral multiplet whose auxiliary component does not vanish on the vacuum. 
The fermion in such chiral multiplet is the goldstino, absorbed by the massive gravitino in the super-Higgs effect, and its complex scalar partner was named sgoldstino in \cite{bfz}. 
In the linear realisations of supersymmetry, the presence of the sgoldstino, with its two real scalar degrees of freedom, is not so trivially reconciled with an acceptable model of large-field inflation. 

Also the scalar partners of quarks and leptons are a cause of concern: their mass terms on the supersymmetry-breaking vacuum are not typically large enough to freeze them during inflation, thus it is crucial to study how they couple to the inflaton in realistic models. 
The same can be said for the extended Higgs boson sector of supersymmetric extensions of the Standard Model, with the additional complication that we should move to a model with a Minkowski vacuum that breaks not only supersymmetry but also the electroweak symmetry: such a discussion goes beyond the aim of the present work. 

\section{Natural sgoldstino-less Minkowski vacua with broken supersymmetry}

After introducing three milestones to be kept in mind in the following, we are almost ready to discuss how close we are to reaching them, and how we can move further. 
Before, however, we need to introduce some background material on no-scale models and on nilpotent chiral superfields. 
After that, we will show how their features can be consistently combined to produce supergravity models with spontaneously broken supersymmetry, naturally vanishing classical vacuum energy and no massless scalar fields in the spectrum.   

\subsection{No-scale models}

As already recalled in the introduction, no-scale models describe supersymmetry breaking with naturally vanishing classical vacuum energy. 
The simplest of such models \cite{noscale} has a single chiral multiplet $T$ and can be defined by 
\begin{equation}
\label{noscaleT}
K = -3 \log (T + \overline{T}) \, , 
\qquad \qquad
W = W_0 \ne 0 \, ,
\end{equation}
where $W_0$ does not depend on $T$ and it is not restrictive to assume that $T + \overline{T} > 0$.
The  K\"ahler potential, parametrising the $SU(1,1)/U(1)$ manifold, satisfies the homogeneity property $K^T K_T = 3$. 
Then, when $W$ does not depend on $T$ as in (\ref{noscaleT}), $F^T F_T = 3 |W_0|^2 / (T + \overline{T})^3 = 3 |W|^2 e^K \ne 0$, the resulting scalar potential is identically vanishing and supersymmetry is spontaneously broken  in Minkowski space all along the complex flat direction $T$. 

For reasons that will be clear momentarily, we rewrite the model in (\ref{noscaleT}) in the new field variable 
\begin{equation}
\label{ztredef}
Z \equiv \frac{2T-1}{2 T+1} \, , 
\qquad \qquad 
\left( T = \frac12 \frac{1+Z}{1-Z} \right) \, ,
\end{equation}
with $|Z| < 1$.
After a K\"ahler transformation, we obtain an equivalent model defined by
\begin{equation}
\label{startZ}
K = - 3 \log(1 - |Z|^2 ) \, , 
\qquad \qquad
W = W_0 \, (1 - Z)^3 \, .
\end{equation}
Since (\ref{ztredef}) is just an analytical field redefinition, the scalar potential is still identically vanishing and supersymmetry is spontaneously broken for any value of $Z$. The advantage of the $Z$ parametrisation of the $SU(1,1)/U(1)$ manifold is that it makes it obvious how to expand $K$ around the self-dual point $Z=0$, corresponding to $T=1/2$ in the $T$ parametrisation. 
Notice that the relative coefficients of the different powers of $Z$ in $W$ are not fine-tuned, they are just the consequence of the classical $SU(1,1)/U(1)$ geometry and of the original assumption that $W$ in (\ref{noscaleT}) does not depend on $T$.  
 
\subsection{Sgoldstino-less models}
\label{sub:nilp}

When local $N=1$ supersymmetry is broken spontaneously, its spin-$1/2$ Nambu--Goldstone fermion, the goldstino, provides the additional degrees of freedom of the massive gravitino. In the absence of vector multiplets, it is not restrictive to assume that the goldstino is the fermionic component of a single chiral goldstino superfield,
\begin{equation}
\label{gold}
S = s + \sqrt{2} \, \theta \, \psi + \theta \theta F^S \, ,
\end{equation}
whose first component $s$ is the complex sgoldstino and whose auxiliary component $F^S$, once properly normalised, is the order parameter of supersymmetry breaking.

In weakly coupled models, sgoldstino masses are related with the curvature of the K\"ahler manifold and are at most of the order of the supersymmetry-breaking scale. 
However, we can also think of strongly coupled models where sgoldstinos are not present in the effective theory and supersymmetry is non-linearly realised in the goldstino sector. 
In global supersymmetry, this corresponds \cite{rocek, linroc, ivakap, CDDFG-nl, komsei} to imposing on the canonically normalised goldstino superfield $S$ the quadratic constraint 
\begin{equation}
\label{S2}
S^2 = 0 \, ,
\end{equation}
which is known to lead, after a field redefinition \cite{kuzenko}, to the original Volkov--Akulov Lagrangian \cite{volaku}.
It was recently shown that such a strategy can be consistently implemented also in supergravity \cite{Farakos:2013ih,ADFS,FKL}, adding a suitable Lagrange multiplier to the original Lagrangian. 
As long as supersymmetry is broken in the $S$ direction, $F^S \neq 0$, the constraint (\ref{S2}) is solved by
\begin{equation}\label{solcon}
s = \frac{\psi \psi}{2 F^S} \, ,
\end{equation}
which implies that $s \psi = s^2 = 0$.
This provides a very simple recipe for the calculation of the scalar potential: we can compute it according to the standard formula (\ref{potential}), as if $S$ were an unconstrained chiral superfield, and then set $\langle s \rangle = 0$.  
As an additional simplification at the superfield level, we can expand $K$ and $W$ in powers of $S$ around $S=0$, setting $S^2 =  0$: after a K\"ahler transformation and a possible constant rescaling, this will lead to a canonical K\"ahler potential and to a superpotential with generic constant and linear terms in $S$. 

\subsection{A sgoldstino-less model with naturally vanishing vacuum energy}
\label{sub:nilpnosc}

We now go back to the single-field no-scale model of (\ref{startZ}) and make the additional assumption that the superfield $Z$ is nilpotent, $Z^2=0$. 
Expanding around $Z = 0$ we get
\begin{equation}
K = 3 |Z|^2 \, , 
\qquad \qquad
W = W_0 \, (1 - 3 Z) \, .
\end{equation}
A simple constant rescaling,
\begin{equation}
Z = - \frac{e^{i \alpha}}{\sqrt3}  \, S \, ,  
\qquad \qquad
(\alpha \in {\mathbb R}) \, ,
\end{equation}
gives then
\begin{equation}
\label{finalW}
K = |S|^2 \, , 
\qquad \qquad
W = W_0 \, (1 + e^{i \alpha} \sqrt3 \, S) \, .
\end{equation}
Since
\begin{equation}
D_S W 
=
\left[
\overline{S} + e^{i \alpha} \sqrt3 \, (1 + |S|^2)\right]\, W_0 \, ,
\end{equation}
we immediately get that, for $\langle s \rangle =0$ as required by (\ref{S2}):
\begin{equation}
\langle F^S \rangle
= 
\langle e^{K/2} \, K^{S \overline{S}} \, \overline{D}_{\overline{S}} \overline{W} \rangle
=
e^{- i \alpha} \sqrt3\, \overline{W_0} \ne 0 \, .
\end{equation}
This implies that, as in the no-scale model considered above, supersymmetry is spontaneously broken with classically vanishing vacuum energy, as in (\ref{minkbro}).
However, in contrast with the no-scale model, the gravitino mass is fixed to $m_{3/2}^2 = |W_0|^2$ and there are no classically flat directions: $S^2=0$ implies $\langle s \rangle =0$ (which corresponds to the self-dual point of $SU(1,1;{\mathbb Z})$ and suggests a possible relation of the nilpotency condition with string-scale supersymmetry-breaking) and the sgoldstino field $s$ is removed from the spectrum.    

Equivalently, we could have replaced $T$ in (\ref{noscaleT}) with $Z=2 T - 1$ and expanded $K$ around  $Z=0$ with the constraint $Z^2=0$.
After a K\"ahler transformation and exploiting again $Z^2=0$, we would have obtained
\begin{equation}
K = \frac34 |Z|^2 \, , 
\qquad
\qquad
W = W_0 \, \left(1 -\frac32 \, Z\right) \, ,
\end{equation}
which is identical to (\ref{finalW}) after identifying $Z = - 2/\sqrt3 \, e^{i \alpha} \, S$.

\subsection{Models with both nilpotent and unconstrained chiral multiplets}
\label{sub:nilpgen}

In the previous subsection we built a sgoldstino-less model with broken supersymmetry and naturally vanishing classical vacuum energy, in terms of a single nilpotent chiral multiplet $S$. We now discuss how to extend this construction to include the additional unconstrained chiral multiplets of a realistic inflationary model, in particular the inflaton multiplet $\Phi$.

Because of the constraint (\ref{S2}), the general form of the K\"ahler potential for the nilpotent goldstino superfield $S$ and $n$ additional unconstrained chiral superfields $\Phi^i$ ($i=1,\ldots,n$) is
\begin{equation}
K = g (\Phi,\overline{\Phi}) + \overline{S} \,  k(\Phi,\overline{\Phi}) + S \, \bar{k} (\Phi,\overline{\Phi}) + |S|^2 \, h(\Phi,\overline{\Phi}) \, ,
\end{equation}
where $k(\Phi,\overline{\Phi})$ can be complex and $g(\Phi,\overline{\Phi})$ and $h(\Phi,\overline{\Phi})$ are real.
Using $\langle s \rangle = 0$, as implied by (\ref{S2}), we can write the inverse K\"ahler metric as 
\begin{equation}
\label{kinv1}
K^{S \overline{S}} = \frac{1}{\widetilde{h}} \, , 
\qquad
\qquad
K^{i \overline{S}} = - \frac{1}{\widetilde{h}} \, g^{i \bar \jmath} \, \bar{k}_{\bar \jmath} \, , 
\qquad
K^{S \bar \jmath} = - \frac{1}{\widetilde{h}} \, k_i \, g^{i \bar \jmath} \, , 
\end{equation}
\begin{equation}
\label{kinv2}
K^{i \bar \jmath} = g^{i \bar \jmath} +  \frac{1}{\widetilde{h}} \, ( g^{i \bar m} \, \bar{k}_{\bar m} ) \,
(  k_l \, g^{l \bar \jmath} ) \, , 
\qquad
\qquad
\qquad
\widetilde{h} = h - k_i \, g^{i\bar \jmath} \bar{k}_{\bar \jmath} \, , 
\end{equation}
where $g_{\bar \imath j} = \partial_{\bar \imath} \, \partial_j \, g $ and $g^{i  \bar \jmath} \, g_{\bar \jmath k} = \delta^i_k$.   
This implies that 
\begin{equation}
F^S = e^{g/2} \, \frac{1}{\widetilde h} \, \left(\overline{D}_{\bar{S}} \overline{W} + k_i \, g^{i \bar \jmath} \, \overline{D}_{\bar \jmath} \overline{W} \right) \, ,
\end{equation}
\begin{equation}
F^i = e^{g/2} \, \frac{1}{\widetilde h} \, \left\{ -  g^{i \bar \jmath} \,  \bar{k}_{\bar \jmath} \, 
\overline{D}_{\bar{S}} \overline{W} + \left[ \widetilde{h} \,  g^{i \bar \jmath} +  
( g^{i \bar m} \, \bar{k}_{\bar m} ) \, (  k_l \, g^{l \bar \jmath} ) \right]
\overline{D}_{\bar \jmath} \overline{W} \right\} \, .
\end{equation}
We see that, if $K$ does not contain non-trivial linear terms in S, $k_i = 0$, the K\"ahler metric is block-diagonal in the nilpotent multiplet $S$  and in the unconstrained multiplets $\Phi^i$:
\begin{equation}
K_{\bar{S} S} = h \, , 
\qquad
K_{\bar{S} j} = K_{\bar \imath S} = 0 \, , 
\qquad
K_{\bar \imath j} =  g_{\bar \imath j} \, .
\end{equation}
We can then extend the model defined in (\ref{finalW}), choosing for definiteness $\alpha=0$ and promoting $W_0$ to an analytic function $f(\Phi)$ of the unconstrained multiplets $\Phi$,
\begin{equation}
W = f (\Phi) \, ( 1 + \sqrt{3} \, S ) \, .
\end{equation}
For $\langle s \rangle =0$, as implied by (\ref{S2}),  it is $\langle F^S F_S \rangle = 3 \, \langle h \, e^{\cal G} \rangle$, thus a sufficient condition to preserve the crucial property $\langle F^S F_S \rangle = 3 \, \langle e^{\cal G} \rangle$ is to have $h=1$. 
Then, if  $F^i F_i = 0$ admits solutions with $\langle f(\phi)\rangle \ne 0$, the potential is positive semi-definite and the vacuum breaks spontaneously supersymmetry in flat space. Moreover, $\langle F^S \rangle \ne 0$ and $\langle F^i \rangle = 0$, consistently with the identification of $S$ with the nilpotent goldstino multiplet. 
Whether all these nice properties continue to hold for a non-trivial $h(\Phi,\overline{\Phi})$ is a model-dependent question.
Away from the vacuum, we may have $F^i \neq 0$, which means that the fermion in $S$ is a mixture of the goldstino and the other fermion fields, but, as long as $f(\phi) \neq 0$ on the whole field space, we can always consistently solve the nilpotency constraint $S^2=0$.

\section{Sgoldstino-less models of inflation with two chiral superfields} 

In the next section, we will present a new class of supergravity models of inflation based on the results of the previous section. 
Before doing that, we critically review some models that have appeared in the recent literature and make use of a nilpotent superfield in combination with another unconstrained superfield~\footnote{There are also `minimal' inflationary models, predating those discussed below, where the inflaton is embedded in the goldstino multiplet \cite{lag1,lag2}, and the latter satisfies the quadratic constraint only in the infrared, after the end of inflation.
These models follow a different approach than ours, since they do not discuss in detail the embedding of the real inflaton in a complex field and the transition from the inflationary background to the vacuum at the end of inflation,
thus we will not discuss them further.}. 
This discussion will allow the reader to compare our new models with the previous ones. 
To facilitate the comparison, we will always label the two chiral multiplets appearing in the models so that $S$ is the goldstino and $\Phi$ the inflaton, and we will always embed the inflaton field $\varphi$ in its superfield $\Phi$ as in (\ref{infdec}).

\subsection{The (F)KL models}
\label{sub:FKL}

A thorough study of supergravity models for inflation with a nilpotent goldstino multiplet and an unconstrained inflaton multiplet has been performed by Ferrara, Kallosh and Linde in \cite{FKL}. 
They propose a class of models with canonical kinetic terms:
\begin{equation}
\label{FKLK}
K = \frac12 (\Phi+ \overline{\Phi})^2 +|S|^2 \, .
\end{equation}
The general form of the superpotential is fixed by the requirement that $S$ be identified with the goldstino superfield for all field configurations relevant during the inflationary phase, i.e. that $F^{\Phi} = 0$ and $F^S \neq 0$ for any value of $\Phi$ scanned during inflation, taking into account the nilpotency constraint.
The result is a superpotential of the form
\begin{equation}
\label{FKLW}
	W = S \, f(\Phi) \,,
\end{equation} 
which is reminiscent of previous models with two unconstrained superfields, where $S$ played the role of the ``stabilizer'' \cite{Kallosh:2010xz, Kallosh:2011qk}.
Understanding again $\langle s \rangle = 0$ in all formulae below, and decomposing $\Phi$ as in (\ref{infdec}), we find
\begin{equation}
F^S = e^{a^2/2} \ \overline{f}(\overline{\phi}) \, ,
\qquad
\qquad
F^\Phi = 0 \, ,
\qquad
\qquad
W = 0 \, ,
\end{equation}
in agreement with the expectations.
The scalar potential is then
\begin{equation}
V = e^{a^2} \, |f(\phi)|^2 \, .
\end{equation}
For appropriate choices of $f$ \cite{Kallosh:2010xz,Kallosh:2011qk}, this allows one to consistently set $a=0$ during inflation and to obtain almost any desired inflationary potential.
However, the requirement of a Minkowski vacuum at the end of inflation implies the existence of a field value for which $f(\phi) = 0$, and this in turn implies that the vacuum is supersymmetric, $\langle F^S \rangle = \langle F^\Phi \rangle =  \langle W \rangle = 0$. 
Besides the fact that unbroken supersymmetry is not what we would eventually ask to a realistic vacuum, at such supersymmetric Minkowski vacua the solution (\ref{solcon}) to the nilpotency constraint becomes singular and the sgoldstino should be restored in the spectrum as an elementary scalar, affecting the field dynamics. 
This fact is not taken into account in the effective supergravity defined by (\ref{FKLK}), (\ref{FKLW}) and (\ref{S2}).

These models were improved by Kallosh and Linde in \cite{KL}, who incorporated supersymmetry breaking by adding a constant term $W_0 \ne 0$ to the superpotential:
\begin{equation}
\label{KLW}
W = S \,  f(\Phi) + W_0 \, .
\end{equation}
The analytic function $f(\Phi)$ should be chosen so that the following conditions are satisfied: 
the potential is minimised at $\phi=0$, with $f(0) \ne 0$ (this also implies $f^{\, \prime} (0) =0$); 
a slow-roll-potential is generated for the inflaton field $\varphi$; 
the scalar partner $a$ of the inflaton is frozen at zero by a large mass term during and after inflation. 
A typical example that satisfies these requirements is $f(\Phi) = M^2 \, (1 - c \, \Phi^2)$, but there are others, the interested reader can find them in \cite{KL}.   
At the minimum, the potential is
\begin{equation}
\label{vminKL}
\langle V \rangle = |f(0)|^2 - 3 \, |W_0|^2 \, , 
\end{equation}
thus to enforce an approximate Minkowski vacuum the function $f(\Phi)$ has to be tuned, to achieve $|f(0)| = \sqrt{3} \, |W_0|$ with a precision of $10^{-60}$. 
Once this is done, however,
\begin{equation}
\langle F^S \rangle = \overline{f}(0) \ne 0 \, , 
\qquad
\langle F^\Phi \rangle = 0 \, , 
\qquad
m_{3/2}^2 = |W_0|^2 \, , 
\end{equation}
and we obtain a self-consistent picture in which supersymmetry is spontaneously broken in Minkowski space, in the direction of the nilpotent goldstino superfield $S$.
The price to pay, however, is the fine-tuning between $f(0)$ and $W_0$.
Note that, in this case, at $\langle s\rangle = 0$
\begin{equation}
	F^\phi = e^{a^2/2} \,\sqrt2 \, a \ \overline{W_0} \, ,
\end{equation}
which vanishes along the inflationary trajectories at $a = 0$.

\subsection{The ADFS model}

This model, proposed by Antoniadis, Dudas, Ferrara  and Sagnotti in \cite{ADFS}, was indeed the first to combine the nilpotent multiplet $S$ with the inflaton multiplet $\Phi$, and is defined by:
\begin{equation}
\label{ADFSmod}
K = -3 \log ( \Phi +\overline{\Phi}  - |S|^2 ) \, , 
\qquad
\qquad
W = W_0 + (f + M \, \Phi)  \, S \, ,
\end{equation}
where $f$ and $M$ are real non-vanishing constants, with $f \, M < 0$.
The constant term $W_0$ in the superpotential can be chosen to be either zero, to reproduce the dual Starobinsky model with a supersymmetric vacuum, or non-zero, to produce supersymmetry breaking at the vacuum.
The K\"ahler potential  in (\ref{ADFSmod}) parametrises the $SU(2,1)/[SU(2) \times U(1)]$ manifold, previously considered \cite{ekn} in a classic example of multi-field no-scale model. 
In contrast with no-scale models, however, the superpotential in (\ref{ADFSmod}) does depend on $\Phi$, a fact that with unconstrained superfields would not allow for a positive semidefinite potential and supersymmetry breaking with naturally vanishing classical vacuum energy. 
However, in the superpotential of (\ref{ADFSmod}) $\Phi$ is multiplied by the nilpotent multiplet $S$, constrained by $S^2=0$, and this is enough to preserve the no-scale properties in the $\Phi$ direction. In fact, $K$ simplifies into 
\begin{equation}
K = -3 \log ( \Phi +\overline{\Phi})  + \frac{3 |S|^2}{\Phi+\overline{\Phi}} \, ,
\end{equation}
which for $\langle s \rangle =0$ (understood in all formulae below) leads to the diagonal K\"ahler metric
\begin{equation}
K_{\bar{S} S} = \frac{3}{\phi+\overline{\phi}} \, , 
\qquad 
K_{\bar{\Phi} \Phi} = \frac{3}{(\phi+\overline{\phi})^{2}} \, , 
\qquad 
K_{\bar{S} \Phi} = K_{\bar{\Phi} S} = 0 \, .
\end{equation}
The scalar potential has then a no-scale structure, 
\begin{equation}
\label{potADFS}
F^\Phi \, F_\Phi  = 3 \, e^{\cal G} = 3 \, e^K \, |W_0|^2 \, , 
\qquad \qquad
F^S \, F_S = \frac{|f + M \phi|^2}{3 \, (\phi + \overline{\phi})^2} \, ,
\end{equation}
and it reduces to the minimal Starobinsky potential for the inflaton $\varphi$, accompanied by a mass term for the axion field $a$:
\begin{equation}
V = \frac{M^2}{12}\left(1 - e^{-\sqrt{\frac23}\varphi}\right)^2 + \frac{M^2}{18}e^{-2\sqrt{\frac23}\varphi} a^2 \, ,
\end{equation}
where
\begin{equation}
\phi =-\frac{f}{M} \, \left( e^{\sqrt{\frac23} \, \varphi} - i \; \sqrt{\frac23} \, a \right)  \, ,
\end{equation}
and
\begin{equation}
{\cal L}_{kin} (\varphi,a) = \frac12 \, (\partial \varphi)^2 + \frac12 \,  e^{-2\sqrt{\frac23}\varphi} \,  (\partial a)^2 \, .
\end{equation}
Note that while $W_0 \ne 0$ modifies some important features of the model, breaking supersymmetry along $\phi$ also at the vacuum, it disappears from the scalar potential.
This model goes a long way towards the three milestones mentioned in the introduction: slow-roll, supersymmetry breaking vacuum for $W_0 \neq 0$, naturally vanishing vacuum energy, large $\varphi$-dependent mass for $a$ during inflation. 
However, there are still some features that may need to be improved.
When $W_0 = 0$ the model has features similar to those of the FKL models in section \ref{sub:FKL}.
For any value of $\phi$ we have $F^\phi = 0$ and $F^s \neq 0$, except at the Minkowski minimum at $\phi = -f/M$, where supersymmetry is restored.
$S$ is therefore identified with the goldstino multiplet along the full inflationary trajectory towards the minimum, which is also a singular point of the solution (\ref{solcon}) to the $S^2 =0$ constraint.
On the other hand, when $W_0 \neq 0$ also $F^\phi \neq 0$ and we cannot interpret $S$ as the goldstino multiplet anymore.
Moreover, also at the supersymmetry breaking vacuum $\langle F^\phi \rangle \neq 0$ and $\langle F^s \rangle = 0$ and supersymmetry is broken in a direction orthogonal to the nilpotent superfield, whose scalar component cannot be identified with a fermion bilinear anymore because (\ref{solcon}) becomes singular.

\subsection{Other models} 
\label{sub:other_models}

Nilpotent superfields have recently been used for building supergravity models of inflation also in \cite{Aoki:2014pna} and \cite{Buchmuller:2014pla}.
The former considers supergravity models with higher derivative terms, thus it explores a rather different direction than the one we are concerned with in this paper.
On the other hand, \cite{Buchmuller:2014pla} analyses supersymmetry breaking mechanisms in the context of inflationary models.
Among others, the authors of   \cite{Buchmuller:2014pla} present an 'O Raifeartaigh model where they add to the K\"ahler potential a very large term proportional to the quartic power of the (canonically normalised) chiral field $S$ whose F-term breaks supersymmetry.
This term decouples the sgoldstino and the resulting model can be effectively described by a nilpotent chiral superfield $S$ coupled to the inflaton $\Phi$ and to the additional chiral field $A$ in the model:
\begin{equation}
	K = \frac12 (\Phi + \overline \Phi)^2 + S \overline S + A \overline A - \xi \, |A|^4 \, , 
	\qquad
	W = S \left(f + \frac12 \,h \, A^2\right) + m\,  A \, \Phi + W_0 \, , 
\end{equation}
where the constants $f$, $m$, $W_0$, $h$, $\xi$ are taken to be real and $S^2=0$.
The structure of $K$ and $W$ resembles the one of the models presented above.
Whenever $|hf| < m^2$, we can have a supersymmetry breaking vacuum at the origin, whose energy is controlled by $\langle V \rangle = f^2-3 \, W_0^2$.
While the consistency of the approach is beyond doubt, once again, the cancellation of the cosmological constant term requires an exquisite fine tuning between $f$ and $W_0$.
Moreover, in this case, during inflation the F-term breaking is not exclusively in the $S$ direction.


\section{A new class of sgoldstino-less models of inflation}

In this section, we propose a class of supergravity models for large-field inflation that extend the single-superfield sgoldstino-less model with naturally vanishing vacuum energy, introduced in subsection~\ref{sub:nilpnosc}, to a two-superfield model containing also the inflaton multiplet. 
They are similar to the models reviewed in Section~3, since they contain a goldstino multiplet $S$, subject to the quadratic constraint $S^2=0$ that removes the sgoldstino degrees of freedom, and an unconstrained inflaton multiplet $\Phi$. 
Also, they allow for slow-roll, large-field inflation, with the scalar partner $a$ of the inflaton field $\varphi$ frozen at $\langle a \rangle = 0$ during the inflationary phase.
However, our models  improve over the models of Section~3 because, thanks to their geometrical properties, they have a supersymmetry breaking vacuum with naturally vanishing cosmological constant at tree level and, at the same time, the direction of supersymmetry breaking at the vacuum is fully aligned with the auxiliary field of the nilpotent goldstino multiplet, consistently with the definition of the latter. 
As we shall discuss in the concluding section, our models can also easily accomodate couplings to a matter sector while preserving their appealing features.

In general, we define our models by 
\begin{equation}
K = \frac12 (\Phi + \bar \Phi)^2 + |S|^2 \, , 
\qquad
\qquad
W = f(\Phi) \, (1 + \sqrt3\, S) \, ,
\end{equation}
where $f(z)$ is a holomorphic function with the following properties:
\begin{equation}
\label{fcond}
\overline{f(z)} = f(- \overline{z}) \, , 
\qquad
\qquad
f^{\, \prime} ( 0 ) = 0 \, , 
\qquad
\qquad
f ( 0 ) \ne 0 \, .
\end{equation}
The first condition means that $f(z)$ can be represented by a power series in $(i \, z)$ with real coefficients, and is analogous to the requirement of a real holomorphic function made in \cite{Kallosh:2010xz, KL}: the difference just reflects the different convention on the embedding of  $\varphi$ in $\Phi$ with respect to (\ref{infdec}).  
The second and third conditions could be relaxed, simply assuming that, for a suitable value of the inflaton field, $\varphi = \varphi_0$, and for a vanishing value of its scalar partner, $a=0$, corresponding to $\phi=\phi_0= i \, \varphi_0 / \sqrt{2}$ in the decomposition (\ref{infdec}), $f^{\, \prime} ( \phi_0 ) = 0$ and $f ( \phi_0 ) \ne 0$.  
If these conditions are satisfied, however, it is always possible to perform an imaginary translation of $\Phi$, corresponding to an analytic field redefinition that leaves the K\"ahler potential invariant, in such a way that $\varphi_0=0$ and therefore (\ref{fcond}) holds.  
We will illustrate such a field redefinition in one of the examples of the next subsection. 

Taking as usual $\langle s \rangle = 0$, as implied by $S^2=0$, then 
\begin{equation}
\label{ourfsfs}
F^S F_S 
= 
3 \, e^{\cal G} 
=
3 \ e^{\frac{\left(\phi+\overline{\phi}\right)^2}{2}} \,  \left| f(\phi) \right|^2 
=
3 \ e^{a^2} \, \left| f  \left( \frac{a + i \, \varphi}{\sqrt{2}} \right) \right|^2 \, ,
\end{equation} 
and the scalar potential is positive definite:
\begin{equation}
\begin{array}{rcl}
V  & = & F^\Phi F_\Phi =
e^{\frac{\left(\phi+\overline{\phi}\right)^2}{2}} \,  \left| f^{\, \prime} (\phi) + f(\phi) \, (\phi + \overline{\phi}) \right|^2 
 \\[3mm]
& = & \displaystyle
e^{a^2} \, \left| f^{\, \prime} \left( \frac{a + i \, \varphi}{\sqrt{2}} \right)
+ \sqrt{2} \, f  \left( \frac{a + i \, \varphi}{\sqrt{2}} \right) \, a \right|^2 \, .
\end{array}
\label{ourV}
\end{equation}
Because of the first condition in (\ref{fcond}), this potential is symmetric under $\phi \to - \overline \phi$, corresponding to $a \rightarrow - a$ and $\varphi \rightarrow \varphi$, thus field configurations with $a=0$ and arbitrary $\varphi$ are always extrema with respect to $a$. 
This also implies that during inflation the scalar $a$ gets a mass of the order of the Hubble parameter, without mass mixing with the inflaton $\varphi$, and is rapidly driven to its VEV, $\langle a \rangle =0$. 
The residual motion of the scalar fields is only in the direction of the inflaton, and the potential driving the inflationary period is then given by 
\begin{equation}
\label{Vinf}
V_{inf}  = \left| f^{\, \prime} \left( \frac{i \, \varphi}{\sqrt{2}} \right) \right|^2 \, .
\end{equation}
At the end of inflation also the inflaton $\varphi$ reaches the minimum of the potential (\ref{ourV}), which, taking into account (\ref{fcond}), is at $\langle \phi \rangle=0$, with:
\begin{equation}
\langle F^S \rangle = \sqrt{3} \, \overline{f} (0) \ne 0 \, , 
\qquad
\langle F^\Phi \rangle = \overline{f}^{\, \prime} (0) = 0 \, , 
\qquad
m_{3/2}^2 = | f (0) |^2 \, .  
\end{equation}
Summarising, we have a Minkowski vacuum that breaks supersymmetry in the direction of the nilpotent goldstino multiplet, as desired. The scalar spectrum around this vacuum is easily computed:
\begin{equation}
\label{masses}
m^2_\varphi= \left| f^{\, \prime \prime} (0) \right|^2 \, , 
\qquad
\qquad
m^2_a =   \left| f^{\, \prime \prime} (0) + 2 \, f(0) \right|^2 \, .
\end{equation}

The inflationary potential of Eq.~(\ref{Vinf}) was also considered in a previous study of models with a single chiral superfield \cite{ket-ter}.
However, there are two important differences with the present study. 
First, $a \rightarrow - a$ is not a symmetry of the full scalar potential of \cite{ket-ter}, thus $\langle a \rangle = 0$ is not a solution of the classical equations of motion and receives corrections. 
Second, at the end of inflation the Minkowski vacua in \cite{ket-ter} have unbroken supersymmetry and, as mentioned in our Introduction, it is far from obvious that supersymmetry breaking in approximately flat space can be achieved by small perturbations of such vacua.   

An important feature of our models is that the inflaton potential (\ref{Vinf}) is controlled by $|f^{\, \prime}|^2$, rather than by $|f|^2$ as in the FKL models of subsection~\ref{sub:FKL} and in the models of \cite{Kallosh:2010ug}, which gives to our models an interesting advantage.  
Our models share with the former the functional freedom in the choice of the inflaton potential.
Given the arbitrary inflationary potential
\begin{equation}
\label{VinfF}
V_{inf} = {\cal F}^2(\varphi) \, , 
\end{equation}
where ${\cal F}(\varphi)$ is a real analytic function of the inflaton, we can reproduce it by choosing
\begin{equation}
\label{intf}
f(\Phi) =-i\, \int d\Phi\, {\cal F}\left( -i\sqrt2\, \Phi\right)\,.
\end{equation}
Notice, however, that $f$ and ${\cal F}$ are defined up to a sign and  that we have the freedom to fix the integration constant to an arbitrary real value $\lambda$, which sets the gravitino mass and the scale of supersymmetry breaking, but disappears from the final inflation potential.

\subsection{Examples} 
\label{sub:examples}

We give below some examples of our new models, leading to potentials that, once restricted to their dependence on $\varphi$ by setting $\langle a \rangle =0$, correspond to the minimal quadratic potential (\ref{linde}), the dual Starobinsky potential (\ref{staro}) and a generalisation of the latter, respectively.

The minimal quadratic inflationary potential (\ref{linde}) is of the form (\ref{VinfF}), for ${\cal F} (\varphi) = (M/\sqrt{2}) \,  \varphi$.
On the basis of the above discussion, it can be derived from our general models by fixing the holomorphic function $f(\Phi)$ according to (\ref{intf}),
\begin{equation}
f(\Phi) = \lambda - \frac{M}{2} \, \Phi^2 \, ,
\end{equation}
where in addition to the real positive parameter $M$ appearing in (\ref{linde}) we have also the real parameter $\lambda$, which can be conveniently chosen to be positive. In this way, $f(\phi)$ is never vanishing during inflation and therefore we are allowed to consistently solve the nilpotency constraint.
It is easy to check that during inflation 
\begin{equation}
m_a^2 (\varphi) >  3 \, M^2 \, \varphi^2 = 6 \, V_{inf} (\varphi) \, . 
\end{equation}
On the vacuum, instead, 
\begin{equation}
m_\varphi^2 = M^2 \, , 
\qquad
\qquad
m_a^2 = (M - 2 \, \lambda)^2 \, , 
\qquad
\qquad
m_{3/2}^2 = \lambda^2 \, .
\end{equation}

The dual Starobinsky potential (\ref{staro}) is also of the form (\ref{VinfF}), with
\begin{equation}
{\cal F} (\varphi) = \sqrt{V_0} \, \left( 1 - e^{- \sqrt{2/3} \, \varphi} \right) \, . 
\end{equation}
We then use (\ref{intf}) and set
\begin{equation}
\label{starof1}
f ( \Phi ) = \lambda
- i \, \sqrt{V_0} \left( \Phi + i \, \frac{\sqrt{3}}{2} \, e^{i \,  \frac{2}{\sqrt{3}} \, \Phi} \right) \, , 
\qquad
(\lambda > 0) \, , 
\end{equation}
which satisfies the conditions (\ref{fcond}) by construction, with $f(0) = \lambda + \sqrt{3 \, V_0} /2$. 
Moreover, also in this case $f(\phi) \neq 0$ for any relevant field value during inflation and once again we can consistently solve the nilpotency constraint.

We could have also derived (\ref{starof1}) starting first from the general ansatz
\begin{equation}
f(\Phi) = \lambda - i \, \mu_1 \Phi + \mu_2\, e^{i\frac{2}{\sqrt3} \Phi} \, ,
\end{equation}
with $\lambda$, $\mu_1$ and $\mu_2$ real positive parameters.
In this case, $f^{\, \prime} (i \varphi/\sqrt{2}) =0$ is solved by
\begin{equation}
\varphi_0 =  - \sqrt{\frac32}\, \log\left(\frac{\sqrt3}{2}\frac{\mu_1}{\mu_2}\right) \, ,
\end{equation}
thus shifting the inflaton field by $\varphi_0$ and setting $\langle a \rangle =0$ we get
\begin{equation}
V_{inf} = \mu_1^2 \, \left(1 - e^{-\sqrt{\frac23}\varphi}\right)^2 \, ,
\end{equation}
which coincides with (\ref{staro}) after the identification $\mu_1^2 = V_0$.

Also in this case, it is easy to check that during inflation 
\begin{equation}
m_a^2 (\varphi) >  6 \, V_{inf} (\varphi) \, . 
\end{equation}
On the vacuum, instead, 
\begin{equation}
m_\varphi^2 = \frac{4}{3}  \, V_0 \, , 
\qquad
m_a^2 = \left( 2 \, \lambda + \sqrt{\frac{V_0}{3}} \right)^2 \, , 
\qquad
m_{3/2}^2 = \left(  \lambda +\frac{\sqrt{3 \, V_0}}{2} \right)^2 \, .
\end{equation}

A generalisation of the inflaton potential (\ref{staro}) considered in \cite{copernicus, Kallosh:2013yoa} is 
\begin{equation}
\label{alfamod}
V_{inf} = V_0 \, \left(1 - e^{-\sqrt{\frac{2}{3 \, \alpha}}\varphi}\right)^2 \, ,
\end{equation}
where $\alpha > 0$ is an additional real parameter. 
The above potential reproduces (\ref{staro}) for $\alpha =1$, and for increasing values of $\alpha$ it gets closer and closer to (\ref{linde}). 
The identification of the function $f(\Phi)$ giving rise to (\ref{alfamod}) is straightforward:
\begin{equation}
\label{alfaf1}
f ( \Phi ) = \lambda
- i \, \sqrt{V_0} \left( \Phi + i \, \frac{\sqrt{3 \, \alpha}}{2} \, e^{i \,  \frac{2}{\sqrt{3 \, \alpha}} \, \Phi} \right) \, , 
\qquad
(\lambda > 0) \, , 
\end{equation}
Once more, we could have derived it starting first from the general ansatz
\begin{equation}
f(\Phi) = \lambda - i \, \mu_1 \Phi + \mu_2\, e^{i\frac{2}{\sqrt{3 \, \alpha}}  \Phi} \, , 
\end{equation}
and performing a shift of the inflaton similar to the one in the previous example. 
Again, during inflation 
\begin{equation}
m_a^2 (\varphi) >  6 \, V_{inf} (\varphi) \, ,
\end{equation}
and on the vacuum 
\begin{equation}
m_\varphi^2 = \frac{4}{3 \, \alpha}  \, V_0 \, , 
\qquad
m_a^2 = \left[ 2 \, \lambda +  \sqrt{\frac{V_0}{3 \, \alpha}} \, (3 \, \alpha - 2 ) \right]^2 \, , 
\qquad
m_{3/2}^2 = \left(  \lambda + \frac{\sqrt{3 \, \alpha \, V_0}}{2} \right)^2 \, .
\end{equation}

\section{Conclusions and outlook}

In summary, we have built supergravity models of large-field inflation, containing a nilpotent goldstino multiplet $S$ and an unconstrained inflaton multiplet $\Phi$, with the following properties: the slow-roll conditions are satisfied in the direction of the inflaton $\varphi$, with its scalar partner $a$ frozen to $\langle a \rangle = 0$ during inflation by a large $\varphi$-dependent mass term; supersymmetry breaking at the end of inflation occurs with naturally vanishing classical vacuum energy and is completely aligned along the goldstino multiplet. 
Although our models represent a non-trivial improvement over the previous ones, many other problems have to be faced on the way towards a realistic model. 

For example, a general problem of inflationary models is how to couple them to the `visible' matter sector of some supersymmetric extension of the Standard Model, without spoiling the nice features of the inflationary potential.
In supergravity it is often assumed that during inflation the scalar fields of the visible sector are frozen at some constant values that do not switch on the corresponding auxiliary fields, and only at the exit from inflation they become dynamical and possibly contribute to the supersymmetry and gauge symmetry breaking mechanisms required in a realistic model.
As already stated in the Introduction, our point of view is that rather than making such an assumption we should test it in concrete models.

As we will now see, our models of Section~4, formulated with just two chiral `hidden-sector' multiplets (goldstino $S$ and inflaton $\Phi$), can be extended without much effort to accommodate additional chiral multiplets in the visible sector, for example squark and slepton multiplets $Z^i$ with $\langle Z^i \rangle = 0$.
We start by assuming a minimal coupling of the matter multiplets $Z^i$ to the inflaton and goldstino multiplets in the K\"ahler potential:
\begin{equation}
K = \frac{(\Phi+\bar \Phi)^2}{2} + |S|^2 + \Delta K (Z,\overline{Z}) \, .
\end{equation}
To preserve the vanishing of the vacuum energy and supersymmetry breaking along the direction of the nilpotent goldstino multiplet, we can promote the superpotential to
\begin{equation}
W = f(\Phi,Z) \, (1 + \sqrt3\, S) \, ,
\end{equation}
where $f(\Phi,Z)$ is a function that does not vanish on the vacuum, $f(0,0) \ne 0$, and such that the potential is invariant under $\phi \to - \bar \phi$.
The potential for $\langle s \rangle =0$ is 
\begin{equation}
V = e^{\Delta K + a^2}\left(|D_{\Phi}f|^2 + |D_i f|^2\right) \, .
\end{equation}
If, for instance, we take canonical kinetic terms for the matter fields, $\Delta K = \sum_i |Z^i|^2$, and the function $f$ contains the $Z^i$ at least quadratically, we see immediately that all the matter fields acquire large masses during inflation, and we can reduce the inflationary potential to 
\begin{equation}
V_{inf} = \left| f^{\, \prime} \left( \frac{i \, \varphi}{\sqrt{2}} ,0 \right) \right|^2 \, .
\end{equation}
At the end of inflation the auxiliary fields of the matter multiplets all vanish, the Minkowski vacuum with broken supersymmetry of our two-field model is preserved and, for canonical $\Delta K$, the supersymmetry-breaking scalar masses are of the order of the gravitino mass. We may also consider the possibility of removing some of the matter scalars from the spectrum, in analogy with what is done for the sgoldstinos. 
In such a case, we would choose to impose $SZ=0$ (rather than $Z^2=0$ as considered in \cite{KL}), along the lines of the corresponding constraint of global supersymmetry \cite{komsei}.   

The discussion becomes more delicate if some of the scalar fields in the visible sector do not appear explicitly in the K\"ahler potential, or if we want them to develop a non-vanishing vacuum expectation value at the end of inflation, as it is the case for the Higgs field: such a discussion goes beyond the aim of the present paper and will be carried out elsewhere. 

A final but important comment concerns the range of validity of supergravity with a nilpotent goldstino multiplet, as dictated by perturbative unitarity.
With complete supermultiplets and exact or spontaneously broken supersymmetry, the ultraviolet cut-off scale of supergravity can be as large as the Planck scale. 
In the presence of a nilpotent goldstino multiplet, however, supersymmetry is non-linearly realised in the supersymmetry-breaking sector. 
In the high-energy limit, $E \gg m_{3/2}$, and on a Minkowski background, gravitino interactions are dominated by the `longitudinal' goldstino components \cite{fayet}, and, restoring for a few lines the explicit powers of $M_P$, four-goldstino amplitudes grow with energy as $s^2/(m_{3/2}^2 M_P^2)$. 
Therefore, on a flat background the effective supergravity theory can be used up to to a cut-off scale of order ${\rm few} \times \sqrt{m_{3/2} M_P}$ (an estimate performed in  \cite{CDDFG-un} finds $s_{cr}=6 \, \sqrt{2 \pi} \, m_{3/2} \, M_P$ for the critical value of the squared centre-of-mass energy), beyond which it becomes strongly interacting. 
The typical energy scale of the classical background in large-field inflation is $E_c \sim V_{inf}^{1/4} (\varphi)$, but the typical energy scale of the quantum fluctuations is $E_q \sim V_{inf}^{1/2} (\varphi)/M_P$.
What is the cut-off scale with which the latter energy should be compared? 
Coming back to units where $M_P=1$, during the inflationary phase the scale of supersymmetry breaking is set by $F^\Phi F_\Phi + F^S F_S$, with $F^\Phi F_\Phi = | f^{\, \prime} (i \, \varphi/ \sqrt{2})|^2 = V_{inf}(\varphi)$ and $F^S F_S = 3 \,  | f (i \, \varphi/ \sqrt{2})|^2 = 3 \, e^G = 3 \, m_{3/2}^2 (\varphi)$.
The non-renormalizable interactions determining the cut-off energy scale during inflation are those associated with the nilpotent goldstino multiplet and controlled by $F^S F_S$, or, equivalently, by the inflaton-dependent gravitino mass $m_{3/2} (\varphi)$.
The condition for perturbative unitarity during inflation is then $E_q^2 \simlt m_{3/2} (\varphi)$, which translates into $ | f^{\, \prime} (i \, \varphi/ \sqrt{2})|^2 \simlt  | f (i \, \varphi/ \sqrt{2})|$.
This condition is essentially insensitive to the scale of supersymmetry breaking on the Minkowski vacuum after the end of inflation, and is comfortably fulfilled in all the examples considered at the end of Section~4.  
We then conclude that sgoldstino-less models of large-field inflation can be simultaneously compatible with perturbative unitarity during the inflationary phase and with weak scale supersymmetry in flat space after the end of inflation, as suggested by the naturalness problem of the Standard Model although not confirmed so far by the LHC.    

\acknowledgments
We thank F.~Bezrukov, S.~Ferrara, M. Pietroni, M.~Porrati, A.W.~Riotto, A.~Sagnotti and M.~Shaposhnikov for discussions. This work was supported in part  by the ERC Advanced Grants no.226455 (\textit{SUPERFIELDS}) and no.267985 (\textit{DaMeSyFla}), by the MIUR grants RBFR10QS5J and PRIN 2010YJ2NYW.

\end{document}